\documentclass[12pt]{article}

\begin{document}

\sloppy
\begin{flushright}{SIT-HEP/TM-10}
\end{flushright}
\vskip 1.5 truecm
\centerline{\large{\bf Kaluza-Klein modes in hybrid inflation}}
\vskip .75 truecm
\centerline{\bf Tomohiro Matsuda
\footnote{matsuda@sit.ac.jp}}
\vskip .4 truecm
\centerline {\it Laboratory of Physics, Saitama Institute of
 Technology,}
\centerline {\it Fusaiji, Okabe-machi, Saitama 369-0293, 
Japan}
\vskip 1. truecm
\makeatletter
\@addtoreset{equation}{section}
\def\theequation{\thesection.\arabic{equation}}
\makeatother
\vskip 1. truecm

\begin{abstract}
\hspace*{\parindent}
When one constructs specific models with the fundamental scale
as low as the TeV scale, there arise many difficulties.
In this paper we examine the hybrid inflation due to bulk
scalar fields, which has been proposed to solve the problem of 
fine-tuning in producing density perturbations.
We find that the Kaluza-Klein modes play
significant roles, which enhance the speed of the phase transition and
 alter the reheating process.
We also argue that a lower bound must be put to the fundamental
 scale, in order to construct the successful hybrid inflation due to the
 bulk scalar fields. 
\end{abstract}

\newpage
\section{Introduction}
\hspace*{\parindent}
In spite of the great success in the quantum field theory, there is still no
consistent scenario in which the quantum gravity is included.
The most promising scenario in this direction would be the string
theory, where the consistency is ensured by the requirement of
additional dimensions.
Initially the sizes of extra dimensions had been assumed to be as small as
$M_p^{-1}$,
but later it has been observed that there is no reason
to believe such a tiny compactification radius\cite{Extra_1}.
The idea of the large extra dimension may solve or weaken the hierarchy
problem. 
Denoting the volume of the $n$-dimensional compact space by $V_n$,
the observed Planck mass is obtained by the relation $M_p^2=M^{n+2}_{*}V_n$,
where $M_{*}$ denotes the fundamental scale of gravity.
If one assumes more than two extra dimensions, $M_{*}$ may be
close to the TeV scale without conflicting any observable
bounds.

Although this new idea inspired creativity in many physicists to lead
them to a new paradigm of phenomenology, a drastic modification is
required for the conventional cosmological scenarios.
Models of inflation and baryogenesis\cite{low-baryon} are especially
sensitive to this low fundamental scale.
To avoid extreme fine tuning, we should reconstruct
the conventional scenarios of the standard cosmology.
This requires inclusion of novel ideas that are quite different from the
conventional one that was used for the models with large fundamental
scale $M_* \sim M_p$, where $M_p$ denotes the Planck scale.
This makes it difficult to construct a specific model for the early
evolution of the Universe.
For example, if one puts the inflaton fields on the brane, their masses
are required to be unnaturally small\cite{fine-tune}.
On the other hand, in generic cases the mass of the inflaton is bounded
from below to achieve successful reheating. 
Thus it seems quite difficult to construct a model for successful
inflation driven by a field on the brane.

A way to avoid these difficulties is put forward by Arkani-Hamed et
al.\cite{Arkani-inflation}, where inflation is assumed to occur 
{\bf before}
the 
stabilization of the internal dimensions.
In this case, however, the late oscillation of the radion is a problem,
which will be solved by the second weak inflation.

One can find other ways to solve problems of inflation in models
with large extra dimensions.
Due to some dynamical mechanisms, the extra dimension may be stabilized
before the Universe exited from inflation.
If the stabilization of the internal dimensions occurred before the
end of inflation, inflation cannot be induced by the fields on the
brane, since their energy densities are highly suppressed.
In this case one may use the bulk field rather than a field on the
brane\cite{Mohapatra-1, Mazumdar-Bulk-Inflation}.

One may find another interesting possibility,  ``brane
inflation''\cite{brane-inflation}, which uses the
interbrane distance as the inflaton.

In this paper we focus our attention to the second possibility,
where the bulk field drives hybrid inflation after the radion
stabilization. 

We first make a brief review of the hybrid bulk field inflation, then 
we will examine the effect of the Kaluza-Klein modes.\footnote{
In ref.\cite{kanti}, interesting aspects of bulk inflation after the
stabilization of the radion field is studied by using
the KK modes in the 4D effective theory. See also ref.\cite{piao}.}
In ref.\cite{Mazumdar-Bulk-Inflation}, it is claimed that the phase
transition becomes so slow that it is
impossible to produce the required density perturbations after inflation
with normal parameter values.
However, we find that the phase transition is fast because of the huge
number of the destabilized Kaluza-Klein modes at the end of inflation.
The number of the destabilized Kaluza-Klein modes becomes up to about
$O(M_p^2/M_*^2)$. 
Although the problem of the slow phase transition seems to be solved by
the Kaluza-Klein modes, another problem is induced
by the excited Kaluza-Klein modes.
The overproduction of the excited Kaluza-Klein modes is the problem,
because they efficiently emit Kaluza-Klein gravitons when they
decay into lower excited modes\cite{Mohapatra-1}.

\section{Hybrid inflation due to bulk scalar fields}
\hspace*{\parindent}
In ref.\cite{Mohapatra-1}, it is argued that a single field inflationary
model cannot provide adequate density perturbations,
either with the inflaton on the brane or with the bulk field
inflation.
Thus it is natural to invoke a hybrid inflationary model in higher
dimensions with the potential for the bulk
field,
\begin{equation}
V(\phi_5,\sigma_5)_{5D}=\lambda^2 M_*
\left(\sigma_{5o}^2-\frac{1}{M_*}\sigma_5^2\right)^2 
+ \frac{m_\phi^2}{2} 
\phi_5^2 
+ \frac{g^2}{M_*} \sigma_5^2 \phi_5^2,
\end{equation}
where $\phi_5$ is the inflaton field with a chaotic initial condition.
The coupling constants $\lambda$ and $g$ are assumed to
be $O(1)$.
To keep things as simple as possible, here we have assumed five
dimensional theory with only one extra dimension.
Extensions to higher dimensional theories are straightforward.
As the higher dimensional fields $\phi_5$ and $\sigma_5$ has a mass of
dimension $3/2$, the interaction terms are nonrenormalizable. 
Higher dimensional fields  $\phi_5,\sigma_5$ are related to the
effective four dimensional fields $\phi, \sigma$ by a scaling
\begin{equation}
\phi=\sqrt{R}\phi_5, \, \sigma=\sqrt{R}\sigma_5,
\end{equation}
which leads to couplings suppressed by the four dimensional Planck mass
$M_p$, where $M_p^2=M_*^{3}R$.
Here $R$ denotes the volume of the extra dimensions.
After dimensional reduction, the effective four dimensional potential
for the 0-modes reads;
\begin{eqnarray}
\label{potential_0}
V(\phi,\sigma)_{4D} &\sim& \lambda^2\frac{M_*^2}{M_p^2}
\left(\frac{M_p^2 \sigma_0^2}{M_*^2} 
-\sigma^2\right)^2\nonumber\\
&&+\frac{m_{\phi^2}}{2}\phi^2 +g^2\left(\frac{M_*}{M_p}\right)^2
\phi^2 \sigma^2.
\end{eqnarray}
Including the Kaluza-Klein modes, the potential becomes\footnote{
We have assumed that the nonrenormalizable terms in the five
dimensional potential (\ref{potential_0}) are obtained by integrating
out massive modes. 
These massive modes, which mediate the interactions $\sigma^4_5$ or
$\phi^2_5\sigma^2_5$, are assumed to have no Kaluza-Klein number.}
\begin{eqnarray}
\label{potential_1}
V(\phi_n, \sigma_n)_{KK} &\sim& \lambda^2\frac{M_*^2}{M_p^2}
\left(\frac{M_p^2 \sigma_0^2}{M_*^2}
-(\sigma^2 + \sum_n \sigma_n^2)\right)^2\nonumber\\
&&+\frac{m_{\phi^2}}{2}(\phi^2 +\sum_n\phi_n^2) 
+g^2\left(\frac{M_*}{M_p}\right)^2
(\phi^2 + \sum_n \phi^2_n) (\sigma^2 + \sum_n \sigma_n^2)\nonumber\\
&&+ \sum_n \frac{n^2}{R^2} \phi_n^2 + \sum_n \frac{n^2}{R^2} \sigma_n.
\end{eqnarray}

Let us first ignore the Kaluza-Klein modes in the effective potential
and discuss the hybrid bulk field inflation induced by the potential
eq.(\ref{potential_0}).
In ref.\cite{Mohapatra-1}, the parameter values $\sigma_o \sim M_* =
10^5$GeV, $m_\phi\sim 10$GeV were considered.
With these values, the density perturbation 
\begin{equation}
\frac{\delta \rho}{\rho} \sim \left(\frac{g}{2\lambda^{3/2}}\right)
\times 10^{-5}
\end{equation}
were obtained.
Then the reheating temperature is calculated within the perturbation
theory.
In ref.\cite{Mohapatra-1}, the decay rates of the inflaton field are
already discussed.
For the decay of the inflaton field into two Higgs fields on the
brane, the decay width becomes
\begin{equation}
\Gamma_{HH} \sim \frac{M_*^4}{32\pi M_p^2 m_\phi}.
\end{equation}
On the other hand, their decay into excited Kaluza-Klein modes are
highly suppressed,
\begin{equation}
\Gamma_{\phi\phi\rightarrow \phi_n \phi_n} 
\sim \lambda^2 \frac{M_*^2}{M_p^2}.
\end{equation}
Thus it seems appropriate to calculate the reheating temperature only 
by the decay into Higgs fields on the brane.
Then the reheating temperature becomes $\sim 100$ MeV, which
satisfies the requirement for the successful big bang nucleosynthesis,
and at the same time solves the problem of overproduction of
Kaluza-Klein gravitons. 
 
In this model, however, a serious problem was reported in
ref.\cite{Mazumdar-Bulk-Inflation}.
The authors of ref.\cite{Mazumdar-Bulk-Inflation} claimed that the phase
transition becomes extremely slow and 
it is impossible to reproduce the present Universe with the above parameter
values. 
Let us first make a brief review of the argument.
The slope of the potential (\ref{potential_0}) in the $\phi$ direction
is given by
\begin{equation}
\label{slope1}
\frac{dV}{d\phi}=\left[2g^2 \left(\frac{M_*}{M_p}\right)^2 \sigma^2 + 
m_{\phi}^2\right]\phi,
\end{equation}
and the inflaton $\phi$ rolls down the potential with the Hubble parameter 
\begin{equation}
H=\sqrt{\frac{8\pi}{3}}\frac{\lambda \sigma_0^2}{M_*}.
\end{equation}
The inflaton $\phi$ slowly rolls down the potential until it reaches the
critical value
\begin{equation}
\phi_c = \frac{\lambda}{g}\left(\frac{\sigma_0M_p}{M_*}\right),
\end{equation}
where the potential in the $\sigma$ direction is destabilized.
The evolution of the inflaton $\phi$ is given by
\begin{equation}
\phi = \phi_c exp\left[-\frac{1}{\sqrt{24\pi}\lambda}
\left(\frac{m_\phi}{\sigma_0}\right)^2 M_* t \right]
\end{equation}
where we take the time variable $t$ as $t=0$ at $\phi=\phi_c$.
In the $\sigma$ direction, the slope of the potential
(\ref{potential_0}) is
\begin{equation}
\label{slope2}
\frac{dV}{d\sigma} = \left(\frac{M_*}{M_p}\right)^2
\left[\lambda^2 \sigma^2 + 2g^2 (\phi^2-\phi^2_c)
\right] \sigma.
\end{equation}
At $t=0$, the initial value of $\sigma$ is expected to be about
$\sigma_{ini} \sim H \sim M_*$.
The phase transition lasts until the first term in eq.(\ref{slope1})
comes to dominate the evolution.
It happens when
\begin{equation}
\label{condition2}
2g^2\left(\frac{M_*}{M_p}\right)^2 \sigma^2\sim m_\phi^2,
\end{equation}
where $\sigma$ becomes about $\sigma_{end} \sim 10^{14}$GeV.
Solving the equation for $\sigma$, one obtains large e-foldings
$N_e \sim 10^5$.
During this period, $\sigma$ evolves from $\sigma_{ini}
\sim 10^5$GeV to $\sigma_{end}\sim 10^{14}$GeV.

However, the Kaluza-Klein modes are important in this case.
Seeing the effective potential eq.(\ref{potential_1}), one can easily
find that not only the 0-mode $\sigma$
but also the Kaluza-Klein modes $\sigma_n$ are destabilized to
contribute the phase  
transition.
The number of the destabilized modes at the end of inflation is
estimated by the potential (\ref{potential_1}), 
\begin{equation}
N_{KK}\sim (R\, \lambda \sigma_0) \sim \left(\frac{M_p}{M_*}\right)^2.
\end{equation}
In the presence of a large number of destabilized channels near $\sigma
\simeq \sigma_{n} \simeq 0$, the phase transition becomes inevitably fast.

In the original calculation of ref.\cite{Mazumdar-Bulk-Inflation},
the initial value of $\sigma$ is $\sigma_{ini}\sim H \sim M_*\sim
10^5$GeV, which is much smaller than the critical value $\sigma_{end}$ 
in eq.(\ref{condition2}).
On the other hand, in the case when $\sigma^2$ in eq.(\ref{slope1})
is replaced by the sum of the huge number of the Kaluza-Klein modes,
the condition (\ref{condition2}) is already satisfied as soon as the
phase transition starts at $\phi\sim \phi_c$.
In this case, each Kaluza-Klein mode $\sigma_n$ may have the initial
value of $\sigma_n \sim H$, if the mass is smaller than the Hubble
parameter. 

Although the problem of the slow phase transition is avoided by the
huge numbers of the Kaluza-Klein modes,
there arises another serious problem.
As is discussed in ref.\cite{Mohapatra-1}, the reheating after the bulk
inflation is not a problem when only the 0-mode oscillates and decays after
inflation.
In this model, however, the decay products of
the excited Kaluza-Klein 
modes will dominate the Universe after inflation,
because the inflation ends with the oscillation and the production of
the excited Kaluza-Klein modes.
Then the Kaluza-Klein gravitons are produced when the excited
Kaluza-Klein modes decay into lower modes.
The decay rate is enhanced by the large number of accessible modes in
the final state, thus the excited modes are very short
lived\cite{Mohapatra-1}.

Let us examine the condition for the Kaluza-Klein gravitons to decay
safe before nucleosynthesis.
The decay width of the Kaluza-Klein gravitons into fields on the brane
is estimated in ref.\cite{kk-graviton-decay},
\begin{equation}
\Gamma \sim \frac{E^3}{M_p^2}.
\end{equation}
Here $E$ denotes the energy of the graviton propagating {\bf in the bulk}.
In the most optimistic case, when $E\sim M_*$, the Kaluza-Klein
gravitons may decay before nucleosynthesis if the fundamental
scale is larger than $10^6$GeV.

\section{Conclusions and Discussions}
\hspace*{\parindent}
In this paper, we have discussed the possibility of successful hybrid
inflation due to the bulk scalar field, in models with large extra
dimensions.
Bulk field inflation is already discussed in papers\cite{Mohapatra-1,
Mazumdar-Bulk-Inflation}, where the problem of Kaluza-Klein
gravitons\cite{Mohapatra-1} and the problem of the slow phase
transition\cite{Mazumdar-Bulk-Inflation} are discussed.
In ref.\cite{Mohapatra-1}, it is argued that the production rate of the 
Kaluza-Klein graviton is so small that the energy of the inflaton is
safely drained into the standard model fields on the brane.
However, including the Kaluza-Klein interactions, the production of the
Kaluza-Klein gravitons becomes efficient.
The problem of the slow phase transition is discussed in
ref.\cite{Mazumdar-Bulk-Inflation}.
However, because of the huge number of excited  Kaluza-Klein states that
becomes unstable at the end of inflation,
the phase transition becomes fast.
The remaining problem is the overproduction of the Kaluza-Klein
gravitons, which puts a lower bound to the fundamental scale.
Even in the most optimistic case, the bound becomes $M_* > 10^6$GeV.

The significant effect of the Kaluza-Klein excited states, which we have
discussed for the bulk inflation, is generically important in
cosmological models that utilizes the phase transition of the bulk field.

\section{Acknowledgment}
We wish to thank K.Shima for encouragement, and our colleagues in
Tokyo University for their kind hospitality.

\end{document}